\documentclass[prb,showpacs,twocolumn,superscriptaddress]{revtex4}
\usepackage{hyperref,graphicx,dcolumn}
\usepackage{amsfonts,amsmath,amssymb,bm,bbm,mathtools,siunitx,mathrsfs}
\usepackage[cal=boondoxo]{mathalfa}
\usepackage{color}
\usepackage{pdfsync}
\usepackage{blkarray}
\usepackage{multirow}
\usepackage{braket}
\usepackage{lipsum}
\usepackage{tikz}
\usepackage{physics}
\usepackage[toc,page]{appendix}

\newcommand{\myFig}[7]{ %
\begin{figure}[htb] 
\begin{center} 
\includegraphics[width=#1\columnwidth,height=#2\columnwidth,clip=true,keepaspectratio=#3,angle=#4]{#5}
\caption{#6} \vspace{-0.5cm} \label{#7} 
\end{center} \end{figure}}

\newcommand{\bigzero}{\mbox{\normalfont\Large\bfseries 0}}
\newcommand{\rvline}{\hspace*{-\arraycolsep}\vline\hspace*{-\arraycolsep}}

\DeclareMathAlphabet{\mathpzc}{OT1}{pzc}{m}{it}
\DeclareMathAlphabet\mathbfcal{OMS}{cmsy}{b}{n}

\begin{document}
\title{On the conservation of the angular momentum in ultrafast spin dynamics}
\author{Jacopo Simoni}\email[Contact email address: ]{jsimoni@lbl.gov}
\affiliation{Molecular Foundry, Lawrence Berkeley National Laboratory, Berkeley, California 94720, USA}
\author{Stefano Sanvito} 
\affiliation{School of Physics, Trinity College, Dublin 2, Ireland}

\begin{abstract}
 The total angular momentum of a close system is a conserved quantity, which should remain constant in time for any excitation experiment once the pumping signal has extinguished. Such conservation, however, is never satisfied in practice in any real-time first principles description of the demagnetization process. Furthermore, there is a growing experimental evidence that the same takes place in experiments. The missing angular momentum is usually associated to lattice vibrations, which are not measured experimentally and are never considered in real-time simulations. Here we critically analyse the issue and conclude that current state-of-the-art simulations violate angular momentum conservation already at the electronic level of description. This shortcoming originates from an oversimplified description of the spin-orbit coupling, which includes atomic contributions but neglects completely that of itinerant electrons. We corroborate our findings with time-dependent simulations using model tight-binding Hamiltonians, and show that indeed such conservation can be re-introduced by an appropriate choice of spin-orbit coupling. The consequences of our findings on recent experiments are also discussed.

\end{abstract}


\maketitle

\section{Introduction}
Thanks to recent advances in femtosecond laser technologies the possibility of achieving control over the magnetization dynamics at time scales of the order of \SI{100}{fs} is now within our reach. After the discovery of the ultrafast optical demagnetization of a Nickel film irradiated with a sub-picosecond laser pulse in 1996\cite{Beaurepaire_1996}, several additional experiments have shown a rich variety of laser-induced phenomena in magnetic compounds, including ultrafast demagnetization\cite{Bigot_2004,Hohlfeld_1997,Gudde_1999,Scholl_1997,Ju_1999,Koopmans_2000,Ogasawara_2005}, spin-reorientation\cite{Kimel_2004,Vomir_2005}, and the modifications of the magnetic structure\cite{Ju_2004,Thiele_2004}. 

Such race towards the control of the magnetization dynamics at the femtosecond time scale is driven by both the fundamental physical investigation of the underlying mechanism leading to ultrafast spin dynamics and technological innovation in the fields of high-speed magnetic recording and spin electronics\cite{Kimel_2005,Stanciu_2007,Berezovsky_2008}.

The now-standard pump-probe experimental protocol represents a valid technique for probing the magnetization dynamics of a sample at short time scales. Here, the system is first excited by the application of an optical pulse (pump), and then the magnetization dynamics is reconstructed by probing it with a second small perturbing signal (probe)\cite{Kimel_2007,Kirilyuk_2010}. Depending on the time delay between the pump and the probe one can accurately trace the magnetization trajectory with a few femtoseconds resolution. Then, the results are typically rationalised by choosing among the most relevant dissipation mechanisms that are at play at the given time scale. In particular, the different types of magnetization dynamics have been usually classified within two main groups.

Fast magnetization processes\cite{Fahnle_2011} are those observed on a time scale ranging between a few nanoseconds to a hundred picoseconds. These are usually described by means of the so-called three-temperatures model, where electrons, phonons and spins form three different thermal baths that are brought out of equilibrium by the application of the laser pulse. The subsystems are able to exchange energy and they thermalise to achieve a final equilibrium state.

An \emph{ultrafast} process, instead, is active at a much shorter time scale, approximately of the order of \SI{100}{fs}. In this case there is much less general agreement on the ultimate cause behind the observed spin dynamics. In recent years different magnetization dynamics models have been proposed in literature, including fully relativistic direct transfer of angular momentum from the laser field to the spins\cite{Hinschberger_2012,Vonesch_2012}, electron-magnon spin-flip scattering\cite{Carpene_2008}, electron-electron spin-flip scattering\cite{Krauss_2009}, Elliott-Yafet mechanism\cite{Carva_2011} and laser induced superdiffusive spin currents\cite{Battiato_2012}. Among all these different schemes only the last one does not require the spin orbit coupling to play a dominant role in the demagnetization process. Crucially, it was experimentally observed that during the ultrafast demagnetization the behaviour of the electronic orbital momentum alone cannot account for the spin decay rate in magnets\cite{Hennecke_2019}. Thus, several theoretical works\cite{Steiauf_2010,Fahnle_2017} suggested that at least part of the spin momentum should be directly transferred to the atomic degrees of freedom. However, the question has not been settled yet, and there is still some debate on the importance of such angular-momentum transfer channel and its relation to the ultrafast demagnetization process.

In our contribution we analyse these issues in great detail, in particular focussing on the conservation of the total electronic angular momentum. We find that the standard treatment of the spin-orbit interaction, based on the sum of atomic contributions, is at the origin of the violation of the electronic angular momentum conservation. Relaxing some of the approximations, and in particular the assumption that in a multi-atom environment the spin-orbit interaction could still be written as if the atoms were isolated, leads to a more complete expression for the spin-orbit interaction. This allows one to satisfy the conservation law. Our results are put to the test with a simple tight-binding model and confirmed by time-dependent simulations.  

In section \ref{sec:2} we will discuss the problem of the total orbital momentum conservation, in particular, with reference to the \emph{ab initio} methods that are more commonly used to simulate these processes. In section \ref{sec:3} we explain why these methods cannot conserve the total system's angular momentum at the present level of development and show how to modify the spin orbit coupling operator in order to enforce this conservation law. In section \ref{sec:4} we look at some results obtained in the case of a very simple tight binding model simulation for atomic clusters and we compare the results with the ones obtained by using the standard models. In section \ref{sec:5} we conclude.  

\section{Non conservation of the total angular momentum in Ab-Initio spin dynamics}\label{sec:2}

The general assumption underpinning any current work in the field of ultrafast magnetism is that the total angular momentum of the system, $\hat{\bf J}$, is conserved during the dynamics. This means that the following equation holds\cite{Fahnle_2013}
\begin{equation}\label{Eq:1}
  \Delta\expval*{\hat{\bf J}} = \Delta\expval*{\hat{\bf S}} + \Delta\expval*{\hat{\bf L}^{\rm e}} + \Delta\expval*{\hat{\bf L}^{\rm atom}} + \Delta\expval*{\hat{\bf L}^{\rm ph}} = 0,
\end{equation}
where $\hat{\bf S}$ is the spin operator, $\hat{\bf L}^{\rm e}$ the electronic orbital momentum operator, $\hat{\bf L}^{\rm atom}$ the atomic orbital momentum and $\hat{\bf L}^{\rm ph}$ the orbital momentum carried by the electromagnetic field interacting with the material. In Eq.~(\ref{Eq:1}) the symbol $\expval*{\hat{\bf O}}=\expval*{\hat{\bf O}}_t$ represents the expectation value of the vector operator $\hat{\bf O}$ at time $t$. Under this assumption the nuclear spin degrees of freedom are not considered, since they do not contribute appreciably to the entire spin dynamics due to the fact that the hyperfine interaction is small compared to the other interactions at play.

When simulating the magnetization dynamics in real time, none of the methods routinely used to interpret the experiments is able to satisfy Eq.~(\ref{Eq:1}). These include \emph{ab initio} approaches, such as real-time time-dependent density functional theory (rtTDDFT)\cite{Krieger_2015,Stamenova_2016}, semi-\emph{ab}-\emph{initio} schemes, such as time-dependent tight-binding (TDTB) models\cite{Tows_2015} or quantum chemistry methods\cite{Chaudhuri_2017}.
In general, this deficiency is not due to a lack of numerical accuracy, instead it represents a well-known limitation of the underpinning theoretical formalism.
The most promising among these methods, both in terms of lack of free parameters and ease of computation, is certainly rtTDDFT\cite{Gross_1990}. In rtTDDFT only the electronic subsystem is evolved quantum mechanically by solving a set of single particle Schr\"odinger-like equations, known as the Kohn-Sham equations\cite{Runge_1984}. By solving exactly the rtTDDFT problem it is possible to reproduce the temporal dependence of both the electron and the magnetization density. However, the interaction between the electrons, the atoms and the laser field are treated by mean of effective external scalar potentials. This means that, in practice, the photons and the atomic degrees of freedom are not self-consistently evolved during the dynamics. As a consequence, only $\Delta\expval*{\hat{\bf S}}+\Delta\expval*{\hat{\bf L}^{\rm e-KS}}$ is accessible from the knowledge of the spin and charge densities, where $\hat{\bf L}^{\rm e-KS}$ is the electron KS angular momentum\cite{comment1}. Unfortunately, as one can deduce from Eq.~(\ref{Eq:1}), $\Delta\expval*{\hat{\bf S}}+\Delta\expval*{\hat{\bf L}^{\rm e-KS}}$ is not a constant of motion.
Hence, if we insist on this level of description, we will not be able to answer to the question on what is the relevant channel for angular momentum dissipation during the demagnetization process. The same considerations are of course valid also for TDTB models, which are based on a similar approach.

In the remaining part of this section we will look at how $\Delta\expval*{\hat{\bf S}}$ and $\Delta\expval*{\hat{\bf L}^{\rm e}}$ evolve in time by solving the set of rtTDDFT equations at the level of implementation provided by the OCTOPUS code\cite{Marques_2003}. OCTOPUS expands the wave function and the operators in real space over a numerical grid and approximates the electron-ion interaction with norm-conserving pseudo potentials\cite{Hamann_1979} incorporating relativistic effects through a spin-orbit coupling term\cite{Kleinman_1980}. In particular, the pseudo potential, $\hat{V}_{\rm ps}(r)$, has the following general form\cite{Oliveira_2008}
\begin{align}\label{Eq:2}
  &\hat{V}_{\rm ps}(r) = \sum_l\sum_{m=-l}^l V_l^{\rm ps}(r)\ket{l,m}\bra{l,m}\,,\nonumber\\
  &V_l^{\rm ps}(r) = \bar{V}_l^{\rm ion}(r)+\frac{V_l^{\rm SO}(r)}{4}+\lambda V_l^{\rm SO}(r)\hat{\bf L}_0\cdot\hat{\bf S}\,,
\end{align}
where $\hat{\bf L}_0$ is the orbital momentum operator of the atom and the expansion is performed over its eigenstates $\ket{l,m}$ (spherical harmonics). In Eq.~(\ref{Eq:2}) the scalar component of the potential, $\bar{V}_l^{\rm ion}(r)$, describes the effect of the mass shift and the Darwin term, while $V_l^{\rm SO}(r)$ sets the range of the spin-orbit coupling that, for convenience, we can rescale by a factor $\lambda$.
\myFig{1}{1}{true}{0}{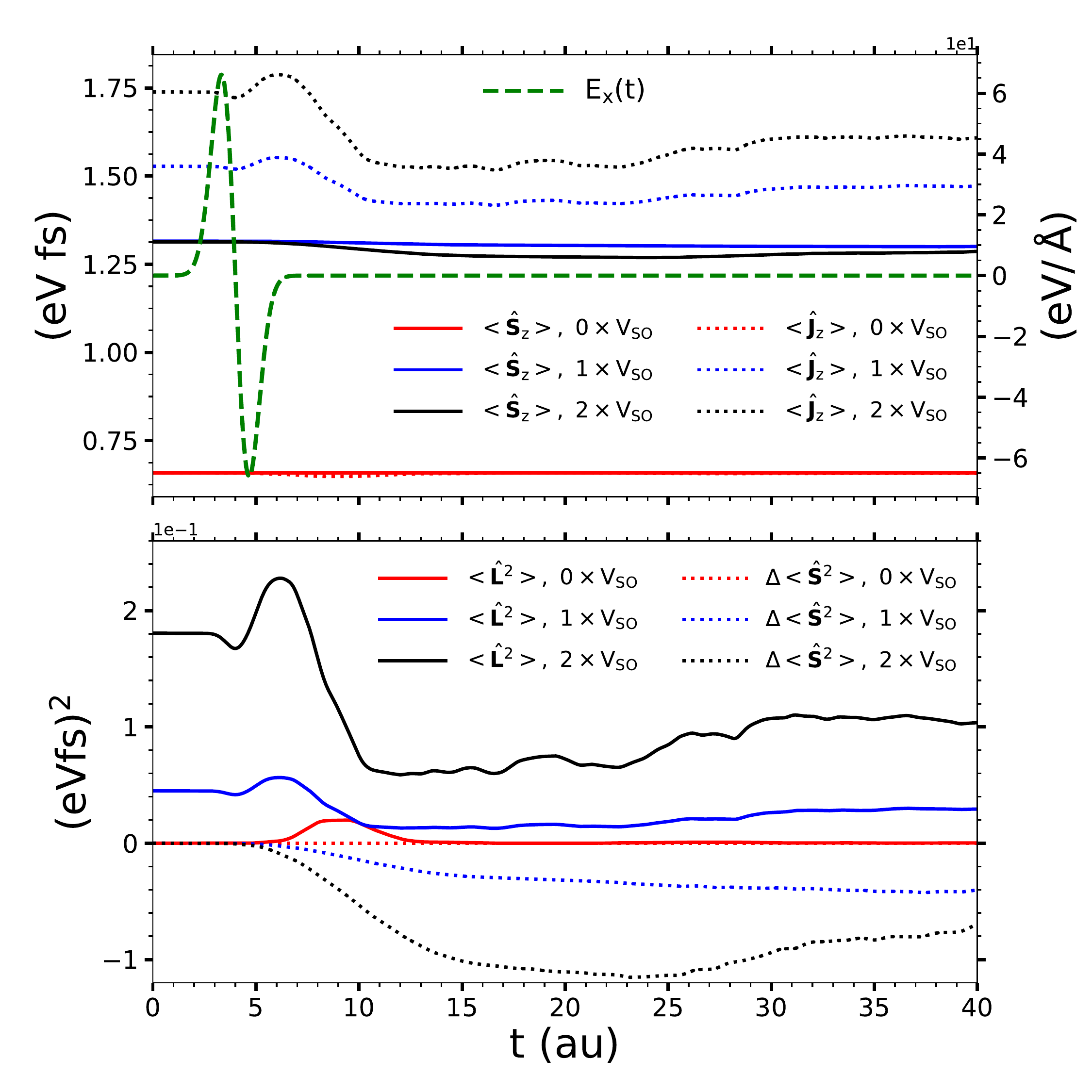}{(Color online) Time evolution of different observables for an isolated Fe atom excited by the application of an external laser field $E_x(t)$ along the x axis (dashed green line, upper panel). The comparison between $\expval*{\hat{S}_z}$ (solid line) and $\expval*{\hat{J}_z}$ (dotted line) is shown in the upper panel for different spin orbit coupling strengths $\lambda V_l^{\rm SO}$ ($\lambda=0, 1, 2$). In the lower panel we show instead $\expval*{\hat{\bf L}^2}$ (solid line) and $\Delta\expval*{\hat{\bf S}^2}$ (dotted line) obtained with the same values of $\lambda$ shown in the upper panel. The units for $\expval*{\hat{S}_z}$ and $\expval*{\hat{J}_z}$ are {\rm eV$\cdot$ fs} (left-hand side scale), while that for the field is {\rm eV/\AA} (right hand-side scale).}{Fig:01}
We start by considering the simplest possible case of an iron atom isolated in vacuum and excited by the application of an external laser pulse. This example, of course, has little practical relevance, but it has important implications for our analysis.
In the case of a single isolated atom, the $\expval*{\hat{{\bf L}}^{\rm atom}}$ term is identically zero, so that, after the electric field has extinguished, the quantity $\expval*{\hat{\bf J}}=\expval*{\hat{\bf S}}+\expval*{\hat{\bf L}^{\rm e}}$ is conserved. This, however, is not the case, as one can see from the time dependent traces of Fig.~(\ref{Fig:01}). We find, in fact, that the total orbital momentum $\expval*{\hat{J}_z}$ still oscillates even after the laser pulse has vanished. Importantly, we notice a difference between the calculations performed with and without spin orbit coupling. If the spin orbit strength, $\lambda$, is set to zero (red lines) $\expval*{\hat{S}_z}$ does not change in time, whereas $\expval*{\hat{J}_z}$ is affected only during the application of the pulse. For $\lambda=1,2$, instead, $\expval*{\hat{J}_z}$ is not conserved even after the application of the pulse with the fluctuations getting larger as the spin-orbit strength increases. The lower panel of Fig.~(\ref{Fig:01}) confirms these findings by showing that $\expval*{\hat{\bf L}_{\rm e}^2}$ remains constant over long times only in the absence of spin-orbit coupling. In contrast, for $\lambda\neq 0$ the module squared of the orbital momentum is not conserved during the evolution, even at times where the laser pulse is not present.
\myFig{1}{1}{true}{0}{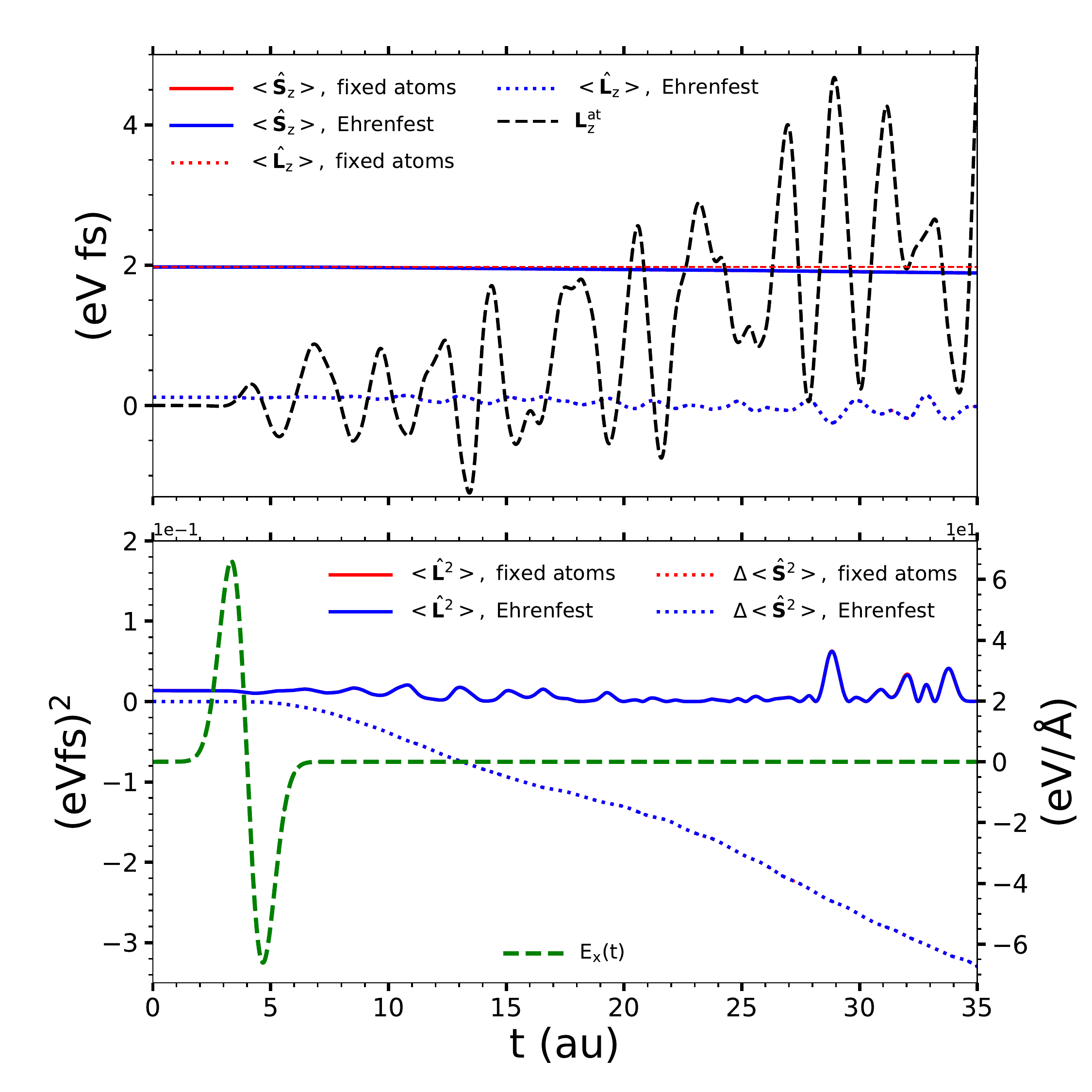}{(Color online) Time evolution of different observables for the iron dimer excited by the application of an external laser field $E_x(t)$ along the x axis (bond axis, dashed green line, lower panel) in presence of spin-orbit coupling. The comparison between $\expval*{\hat{S}_z}$ (solid line) and $\expval*{\hat{L}_z^{\rm e}}$ (dotted line) is shown in the upper panel. It is obtained by fixing the atomic positions during the dynamics and by performing a Ehrenfest molecular dynamics calculation. In this last case also the total atomic orbital momentum is shown (dashed black line). In the lower panel we show instead $\expval*{\hat{\bf L}^2_{\rm e}}$ (solid line) and $\Delta\expval*{\hat{\bf S}^2}$ (dotted line) obtained for \emph{fixed atoms} and for Ehrenfest dynamics calculations. The units for $\expval*{\hat{S}_z}$ and $\expval*{\hat{J}_z}$ are {\rm eV$\cdot$ fs} (left-hand side scale), while that for the field is {\rm eV/\AA} (right-hand side scale).}{Fig:02}
The next level of complexity is achieved by looking at a ${\rm Fe_2}$ ferromagnetic dimer. In this case we first obtain the electronic ground state at the optimized geometry and then we let the system evolve in time under the effect of the same laser field of Fig.~(\ref{Fig:01}), polarized along the bond axis of the dimer. In the simulation the atomic degrees of freedom are explicitly included by performing the rtTDDFT evolution together with Ehrenfest molecular dynamics for the ions\cite{Andrade_2009}
In Fig.~(\ref{Fig:02}) we compare the temporal evolution obtained without allowing the atomic motion to that obtained from Ehrenfest dynamics. In the upper panel we look at $\expval*{\hat{L}_z^{\rm e}}$ and $\expval*{\hat{S}_z}$ and we note that the atomic motion plays no role in their dynamics (the curves obtained with Ehrenfest dynamics overlap perfectly with the ones obtained by keeping the atoms fixed). The time scale is, in fact, too short and the atoms do not have enough time to move. In the case of fixed atomic coordinates the quantity $\expval*{\hat{L}_z^{\rm e}}+\expval*{\hat{S}_z}$ is not conserved, as expected. However, also the quantity $\expval*{\hat{L}_z^{\rm e}}+\expval*{\hat{S}_z}+L_z^{\rm atom}$ is not conserved during the Ehrenfest dynamics. In particular, the atomic orbital momentum is characterized by huge oscillations persisting also after that the pulse amplitude sets to zero. In the lower panel of Fig.~(\ref{Fig:02}) it is confirmed that the spin lost during the dynamics is not transfered to the orbital degrees of freedom. The same conclusions are valid for both atomic clusters and bulk systems, and increasing the size of the cluster does not help to reinstate angular momentum conservation (see also Ref.~\onlinecite{Stamenova_2016}).

In the next section we will critically review the approximations taken in our time-dependent simulations and show that the explicit form of spin-orbit potential used is the source of the non-conservation of the angular momentum. Such oversimplified description of the spin-orbit coupling can be corrected and the conservation re-instated. 
 
\section{Breakdown of the conservation law and solution}\label{sec:3}
For the remaining of the paper we will focus on finite size systems, however, these arguments can be generalized also to the case of extended, periodic, systems.
\subsection{case 1) $\hat{H}_{\rm SO}=0$}
The first case we analyse corresponds to the situation where there is no spin-orbit interaction. As a consequence, our attention is focussed on the conservation of the total angular momentum only (we do not consider the spin, given that it commutes with the system's Hamiltonian, it is a constant of motion).
We start by writing the Hamiltonian $\hat{H}$ of a system of $N_e$ electrons in the field generated by $N_i$ identical ions of mass $M$ (we make this choice to keep the treatment easier). Here the ions are treated also as quantum particles, there is no additional external field. The Hamiltonian thus writes,
\begin{equation} \label{Eq:H}
  \hat{H} = \sum_{a=1}^{N_i}\frac{\hat{\bf P}_a^2}{2M} + \hat{V}_{\rm ii} + \hat{H}_{\rm ie}(\hat{\bf r},\hat{\bf p},\hat{\bf R})\,,
\end{equation}
where $(\hat{\bf r},\hat{\bf p})$ defines the set of electronic positions and linear momentum operators, $\hat{\bf R}=(\hat{R}_{1,x},\hat{R}_{1,y},\ldots\hat{R}_{N_i,z})$ is the set of $3N_i$ atomic coordinate operators, $\hat{V}_{\rm ii}$ is the ion-ion interaction and $\hat{H}_{\rm ie}$ represents the full electronic Hamiltonian. This latter includes the electron kinetic energy term, the electron-electron interaction and the electron-ion interaction. By using $d\hat{\bf R}_a/dt=\hat{\bf P}_a/M$, it is easy to show that the time derivative of the orbital momentum associated to ion $a$ is
\begin{equation}
\frac{d}{dt}\hat{\bf L}_a = -\hat{\bf R}_a\times\nabla_{{\bf R}_a}\Big[\hat{H}_{\rm ie}+\hat{V}_{\rm ii}\Big]\,,
\end{equation}
by writing $\hat{H}_{\rm ie}=\hat{H}_{\rm ie}^0+\hat{U}_{\rm ee}$, where $\hat{U}_{\rm ee}$ is the electron-electron interaction potential and $\hat{H}_{\rm ie}^0=\sum_{\rm i=1}^{N_e}\hat{\bf p}_i^2/2m_e+\sum_{\rm i=1}^{N_e}\sum_{a=1}^{N_i}v_{\rm ie}(|\hat{{\bf r}}_i-\hat{{\bf R}}_a|)$, with $v_{\rm ie}(|\hat{\bf r}_{\rm i}-\hat{\bf R}_a|)$ being the electrostatic potential exerted by the $a$-th ion on the {\rm i}-th electron, it is easy to show that
\begin{align}\label{Eq:dLadt}
  &\frac{d}{dt}\hat{\bf L}_a = -\hat{\bf R}_a\times\bigg[\sum_{i=1}^{N_e}\nabla_{{\bf R}_a}v_{\rm ie}(|\hat{\bf r}_i-\hat{\bf R}_a|) + \nabla_{{\bf R}_a}\hat{V}_{\rm ii}\bigg]\,,\nonumber\\
&=-\sum_{i=1}^{N_e}\hat{\bf r}_i\times\nabla_{{\bf R}_a}v_{\rm ie}(|\hat{\bf r}_i-\hat{\bf R}_a|) - \hat{\bf R}_a\times\nabla_{{\bf R}_a}\hat{V}_{\rm ii}\,,
\end{align}
%
%
%
by combining Eq.~(\ref{Eq:dLadt}) with the time derivative of the ionic linear momentum operator, $d\hat{\bf P}_a/dt = -\nabla_{{\bf R}_a}\big[\hat{H}_{\rm ie}+\hat{V}_{\rm ii}\big]$, we obtain the following important relation
\begin{align} \label{Eq:Lconserv}
&\frac{d}{dt}\hat{\mathbfcal{L}}_a =\frac{d}{dt}\bigg\{\hat{\bf L}_a - \sum_{i=1}^{N_e}\hat{\bf r}_i\times\hat{\bf P}_a\bigg\} =\nonumber\\
&= -\frac{1}{m_e}\sum_{i=1}^{N_e}\hat{\bf p}_i\times\hat{\bf P}_a+\sum_{i=1}^{N_e}\hat{\bf r}_i\times\nabla_{{\bf R}_a}\hat{V}_{\rm ii}-\hat{\bf R}_a\times\nabla_{{\bf R}_a}\hat{V}_{\rm ii}\,,
\end{align}
from which we can conclude that $\sum_{a=1}^{N_i}\hat{\mathbfcal{L}}_a$ is a constant of motion for the system's Hamiltonian $\hat{H}$,
\begin{equation}\label{Eq:Ltconserv}
\frac{d}{dt}\sum_{a=1}^{N_i}\hat{\mathbfcal{L}}_a=0\,,
\end{equation}
(further details are given in \ref{app:1}). In addition, we can rewrite $\sum_{a=1}^{N_i}\hat{\mathbfcal{L}}_a=\hat{\bf L}^{\rm atom}+\hat{\bf L}^{\rm e}$ with Eq.~(\ref{Eq:1}) being manifestly satisfied in the case of a spin unpolarized system with no externally applied electromagnetic field. It may be also useful to separate the electronic contribution to the atomic angular momentum from the ionic one, by defining $\hat{\mathbfcal{L}}_a^{\rm e}=-\sum_{i=1}^{N_e}\hat{\bf r}_i\times\hat{\bf P}_a$ we can finally write $\hat{\mathbfcal{L}}_a=\hat{\bf L}_a+\hat{\mathbfcal{L}}_a^{\rm e}$.
\paragraph{Proof of the inequality $\hat{\mathbfcal{L}}_a^{\rm e}\neq \hat{\bf L}_0^a:$\\} 
Let us define the angular momentum $\hat{\bf L}_a^0$ as that associated to the atom $a$ in the absence of other nuclei, this operator commutes with the Hamiltonian of the isolated atom: $\hat{H}_{\rm ie}^a=\sum_{i=1}^{N_e}\hat{\bf p}_i^2/2m_{\rm e}+\sum_{i=1}^{N_e}v_{\rm ie}(|\hat{\bf r}_i-\hat{\bf R}_a|)+\hat{U}_{\rm ee}$.
In general it is easy to show that for every multi-atom system the inequality $[\sum_{a=1}^{N_i}\hat{\bf L}_a^0,\hat{H}_{\rm ie}]\neq 0$ is valid due to the broken spherical symmetry of the system (namely to the term $\sum_{i=1}^{N_e}\sum_{b\neq a}^{N_i}v_{\rm ie}(|\hat{\bf r}_i-\hat{\bf R}_b|)$).
 
From the previous considerations we have also $[\sum_{a=1}^{N_i}\hat{\mathbfcal{L}}_a,\hat{H}_{\rm ie}+\hat{V}_{\rm ii}]=0$, while for the purely electronic part we write $-i[\hat{\mathbfcal{L}}_a^{\rm e},\hat{H}_{\rm ie}]/\hbar=\hat{{\bf R}}_a\times\nabla_{{\bf R}_a}\hat{H}_{\rm ie}$. In general it can be shown that $-i[\hat{\bf{L}}_a^0,\hat{H}_{\rm ie}]/\hbar\neq\hat{{\bf R}}_a\times\nabla_{{\bf R}_a}\hat{H}_{\rm ie}$ (all the details are given in \ref{app:B}) and, as a consequence, $\hat{\mathbfcal{L}}_a^{\rm e}\neq\hat{\bf L}_a^0$. In conclusion the electronic orbital momentum operator corresponding to $\hat{\bf L}^{\rm e}=\sum_{a=1}^{N_i}\hat{\mathbfcal{L}}_a^{\rm e}$ cannot be identified with the sum of the operators $\hat{\bf L}_a^0$ for isolated atoms. 
\paragraph{Isolated atom case}
In the case of a single isolated atom we have $[\hat{\bf L}_a^0,\hat{H}_{\rm ie}^a]=0$ due to spherical symmetry, in addition $\nabla_{{\bf R}_a}\hat{H}_{\rm ie}^a=0$, and, as a consequence $\hat{\mathbfcal{L}}_a^{\rm e}=\hat{\bf L}_a^0+{\bf c}\hat{\mathbb{I}}$ where we can freely set ${\bf c}=0$.
\paragraph{Non conservation of $\hat{\bf L}_{\rm at}+\hat{\bf L}_{\rm elec}$ in TDTB\\}
In TDTB models the electronic orbital momentum operator around each atom is usually set by definition to $\hat{\mathbfcal{L}}_a^{\rm e}=\hat{\bf L}_a^0$. The time derivative of the expectation value of the electronic orbital momentum operator around atom $a$ is (see \ref{app:B})
\begin{align} \label{TDTB}
\frac{d}{dt}{\bf L}_a^0&=-\frac{i}{\hbar}\Big\langle\Big[\hat{\bf L}_a^0,\hat{H}_{\rm ie}+\hat{V}_{ii}\Big]\Big\rangle=-\frac{i}{\hbar}\Big\langle\Big[\hat{\bf L}_a^0,\hat{H}_{\rm ie}\Big]\Big\rangle\nonumber\\
&\neq{\bf R}_a\times<\nabla_{{\bf R}_a}\hat{H}_{\rm ie}>+{\bf R}_a\times\nabla_{{\bf R}_a}V_{ii}=-\frac{d}{dt}{\bf L}^{\rm atom}_a\,,
\end{align}
that leads to
\begin{equation}
\frac{d}{dt}\sum_{a=1}^{N_{\rm i}}{\bf L}_a^0 + \frac{d}{dt}{\bf L}^{\rm atom} \neq 0\,.
\end{equation}
As a consequence these models cannot conserve the total orbital momentum even, if the atoms are allowed to move. Note that This result is independent on the choice of $\hat{H}_{\rm ie}$ while $\hat{U}_{\rm ee}$ is usually parametrized by means of a mean field approximation.\cite{Tows_2015}
\subsection{case 2) $\hat{H}_{\rm SO}\neq 0$}
Here we consider the case of finite spin orbit coupling, so that the spin degrees of freedom need to be re-introduced in the discussion. We will not generalize the treatment to the full Dirac formalism, since it is unnecessary for our conclusions, and so we will simply add the spin-orbit coupling term to the general Hamiltonian of Eq.~(\ref{Eq:H}). The spin-orbit coupling operator associated to atom $a$ may be written in the following general form
\begin{equation}
\hat{H}_a^{\rm SO} = -\frac{e\hbar}{4m_e^2 c^2}\sum_{i=1}^{N_e}\hat{\boldsymbol{\sigma}}_i\cdot\big[\tilde{\bf E}_a(\hat{\mathbf{r}}_i,t)\times\hat{\bf p}_i\big]\,,
\end{equation}
where $\tilde{\bf E}_a(\hat{{\bf r}},t)$ is the effective screened electric field due to atom $a$ and experienced by the electrons. From now on we will treat the electron-electron interaction in a mean-field way, so that the expression for the effective field can be written as, $\tilde{\bf E}_a(\hat{\bf r},t)={\bf E}_{\rm ext}(t)/N_i+\tilde{\bf E}^{\rm mf}_a(\hat{\bf r}-\hat{\bf R}_a)$, where $\tilde{\bf E}^{\rm mf}_a$ is short range due to the screening of the conduction electrons, while ${\bf E}_{\rm ext}$ is the applied external field assumed to be spatially homogeneous. By summing over all the atoms the spin-orbit Hamiltonian becomes
\begin{equation}
\hat{H}^{\rm SO} = \hat{H}^{\rm SO-ext} + \sum_{a=1}^{N_{\rm i}}\hat{\tilde{H}}_a^{\rm SO}\,,\qquad\textrm{where,}
\end{equation}
\begin{align}
&\hat{H}^{\rm SO-ext} = -\frac{e\hbar}{4m_e^2 c^2}\bigg[{\bf E}_{\rm ext}(t)\cdot\sum_{i=1}^{N_{\rm e}}\hat{\bf p}_i\times\hat{\boldsymbol{\sigma}}_i\bigg]\nonumber\,,\\
&\hat{\tilde{H}}_a^{\rm SO} = -\frac{e\hbar}{4m_e^2 c^2}\sum_{i=1}^{N_{\rm e}}\hat{\boldsymbol{\sigma}}_i\cdot\Big[\tilde{\bf E}_a^{\rm mf}(\hat{\bf r}_i-\hat{\bf R}_a)\times\hat{\bf p}_i\Big]\,.
\end{align}
The first contribution due to the externally applied electric field is not of great interest here since it will act only during the application of the laser pulse, while we are mainly interested in the long term dynamics of the system, therefore, we will focus on the second term only. Without any loss of generality we can rewrite the mean-field screened electric field as $e\tilde{\bf E}^{\rm mf}_a=-\nabla_{{\bf R}_a}\tilde{v}_{\rm ie}(|{\bf r}-{\bf R}_a|)$, where $\tilde{v}_{\rm ie}$ is the screened electron-ion potential. The spin-orbit coupling operator then becomes
\begin{align}
&\hat{H}_{\rm SO}^a = \frac{i}{2m_e^2 c^2\hbar}\sum_{i=1}^{N_e}\hat{\bf S}_i\cdot\bigg\{\big[\hat{\bf P}_a,\tilde{v}_{\rm ie}(|\hat{\bf r}_i-\hat{\bf R}_a|)\big]\times\hat{\bf p}_i\bigg\}\nonumber\\
&=\frac{1}{2m_e^2 c^2}\sum_{i=1}^{N_e}\sum_{ljk}\epsilon_{ljk}\hat{S}_i^l\hat{P}_a^j\bigg\{-\frac{i}{\hbar}\big[\hat{p}_i^k,\tilde{v}_{\rm ie}(|{\bf r}-{\bf R}_a|)\big]\bigg\}+\nonumber\\
&+\frac{i}{2m_e^2 c^2\hbar}\sum_{i=1}^{N_e}\sum_{ljk}\epsilon_{ljk}\hat{S}_i^l\bigg[\hat{P}_a^j\hat{p}_i^k,\tilde{v}_{\rm ie}(|\hat{\bf r}_i-\hat{\bf R}_a|)\bigg]\,.
\end{align}
The second term is exactly zero since the electronic and atomic momentum commute. Thus, the spin-orbit coupling experienced by a single electron becomes
\begin{align} \label{Eq:soc}
\hat{H}_{\rm SO}^a &= -\frac{\partial_r\tilde{v}_{\rm ie}(r)}{2 m_e^2 c^2|\hat{\bf r}-\hat{\bf R}_a|}\hat{\bf S}\cdot\big\{(\hat{\bf R}_a-\hat{\bf r})\times\hat{\bf P}_a\big\}\nonumber\\
&= -\frac{\partial_r\tilde{v}_{\rm ie}(r)}{2 m_e^2 c^2|\hat{\bf r}-\hat{\bf R}_a|}\hat{\bf S}\cdot\hat{\mathbfcal{L}}_a\,
\end{align}
where $\hat{\mathbfcal{L}}_a$ is the orbital momentum operator introduced in Eq.~(\ref{Eq:Lconserv}). It is evident that $\sum_{a=1}^{N_i}\mathbfcal{L}_a$ is not a constant of motion of the new Hamiltonian $\hat{H}'=\hat{H}+\hat{H}_{\rm SO}$, with $\hat{H}$ given by Eq.~(\ref{Eq:H}). In particular from Eq.~(\ref{Eq:soc}), by writing $\hat{H}_{\rm SO}=\sum_af_a({\bf r})\hat{\mathbfcal{L}}_a\cdot\hat{\bf S}$, it is easy to show that
\begin{align}
&[\hat{\mathbfcal{L}}_a,\hat{H}_{\rm SO}] = (\hat{\bf R}_a-\hat{\bf r})\times\sum_{b=1}^{N_i}\big[\hat{\bf P}_a,f_b(\hat{{\bf r}})\big]\hat{\mathbfcal{L}}_b\cdot\hat{\bf S}+\nonumber\\
&+\sum_{b=1}^{N_i}f_b(\hat{\bf r})\big[\hat{\mathbfcal{L}}_a,\hat{\mathbfcal{L}}_b\big]\cdot\hat{\bf S}\,,
\end{align}
where the first term on the right hand side is exactly zero. Furthermore, given the fact that $\hat{\mathbfcal{L}}_a$ is a generator of the spatial rotation group and satisfies the commutation relations $[\hat{\mathbfcal{L}}_{a,x},\hat{\mathbfcal{L}}_{b,y}]=i\hbar\delta_{a,b}\sum_z\varepsilon_{xyz}\hat{\mathbfcal{L}}_{a,z}$, we easily obtain $d\hat{\mathbfcal{L}}_a/dt=f_a({\bf r})\hat{\bf S}\times\hat{\mathbfcal{L}}_a$. Since $d\hat{\bf S}/dt=\sum_{a=1}^{N_i}f_a({\bf r})\hat{\mathbfcal{L}}_a\times\hat{\bf S}$ we conclude that $d\hat{\bf J}/dt=0$ with $\hat{\bf J}=\sum_{a=1}^{N_i}\hat{\mathbfcal{L}}_a+\hat{\bf S}$.
\paragraph{Non conservation of the orbital momentum in non collinear TDDFT}
The non conservation of the total angular momentum in the rtTDDFT simulations presented in Figs.~(\ref{Fig:01} and \ref{Fig:02}) and reported previously in reference~\onlinecite{Stamenova_2016} is due to the fact that the system's spin orbit coupling operator (introduced via the pseudopotential approximation) is given by $\hat{H}_{\rm SO}({\bf r})=\sum_{a=1}^{N_i}\hat{V}_{\rm ps}^{\rm SO}({\bf r}-{\bf R}_a)$ and it is a function of the isolated atom orbital momentum operator $\hat{\bf L}_0$. As a consequence $d\hat{\mathbfcal{L}}_a/dt\neq f_a({\bf r})\hat{\bf S}\times\hat{\mathbfcal{L}}_a$ and $\hat{\bf J}$ is not a constant of motion. Note that also $\sum_{a=1}^{N_i}\hat{\bf L}_a^0+\hat{\bf S}$ is not a constant of motion, since $\sum_{a=1}^{N_i}\hat{\bf L}_a^0$ does not commute with the crystal field Hamiltonian $\hat{H}_{\rm ie}^0$.
\section{Results}\label{sec:4}
In this section we consider a set of different atomic clusters and we analyze their temporal evolution by using the correct orbital-momentum-conserving spin-orbit coupling [see Eq.~\ref{Eq:soc}]. The results are then compared with the approximated, non-orbital-momentum-conserving evolution.
For the analysis we use a TB approximation, whose Hamiltonian writes as follows
\begin{align}
&\hat{H}(t) = \sum_{a=1}^{N_i}\frac{{\bf P}_a^2}{2M}+\hat{\tilde{H}}_{\rm ie}^{[{\bf R}]}(t) + V_{\rm ii}({\bf R}) + \sum_{a=1}^{N_i}Z_a^* e{\bf R}_a\cdot{\bf E}(t),\\
&\hat{\tilde{H}}_{\rm ie}^{[{\bf R}]}(t) = \sum_{a=1}^{N_i}\sum_i\epsilon_{ai}\hat{c}^\dagger_{ai}\hat{c}_{ai} - \sum_{a<b}\sum_{i<j}[t_{ab}^{ij}({\bf R})\hat{c}^\dagger_{bj}\hat{c}_{ai} + {\rm h.c.}] +\nonumber\\
&+ \hat{H}_{\rm SO}^{[{\bf R}]} + \hat{\bf D}\cdot{\bf E}(t)\label{Eq:hamilt}\,.
\end{align}
In the electronic Hamiltonian, $\hat{\tilde{H}}_{\rm ie}^{[{\bf R}]}$, the first two terms on the right hand side correspond to $\hat{H}^0_{\rm ie}$, namely to the electron-ion interaction and the electronic kinetic energy (this term can also eventually include a mean-field expression for the electron-electron interaction, not considered here). The last term on the right hand side is the electric dipole interaction ($\hat{\bf D}=-e\hat{\bf r}$). We have further defined $Z_a^*$ as the effective atomic number and ${\bf E}(t)$ as the externally applied electric field.
\subsection{Comparison between the two types of evolution}
Here we compare the temporal evolution of an atomic dimer under the effect of the two different spin orbit coupling operators introduced in the previous section. The basis set is built of the $s$ and $p$ orbitals of the two atoms. The model here for explanatory reasons does not have to be physically realistic and we will focus on more realistic systems in future works. In addition the model depends on a set of free parameters that are set arbitrarily and varied in order to look at their influence on the dynamics. These are hopping terms $t_{ss}$, $t_{sp}$ and $t_{pp}$, the spin orbit strengths $\lambda_a$, the on-site energies $\epsilon_{ai}$ and the electric dipole parameters. The atoms are assumed from now on to be identical.
\myFig{1}{1}{true}{0}{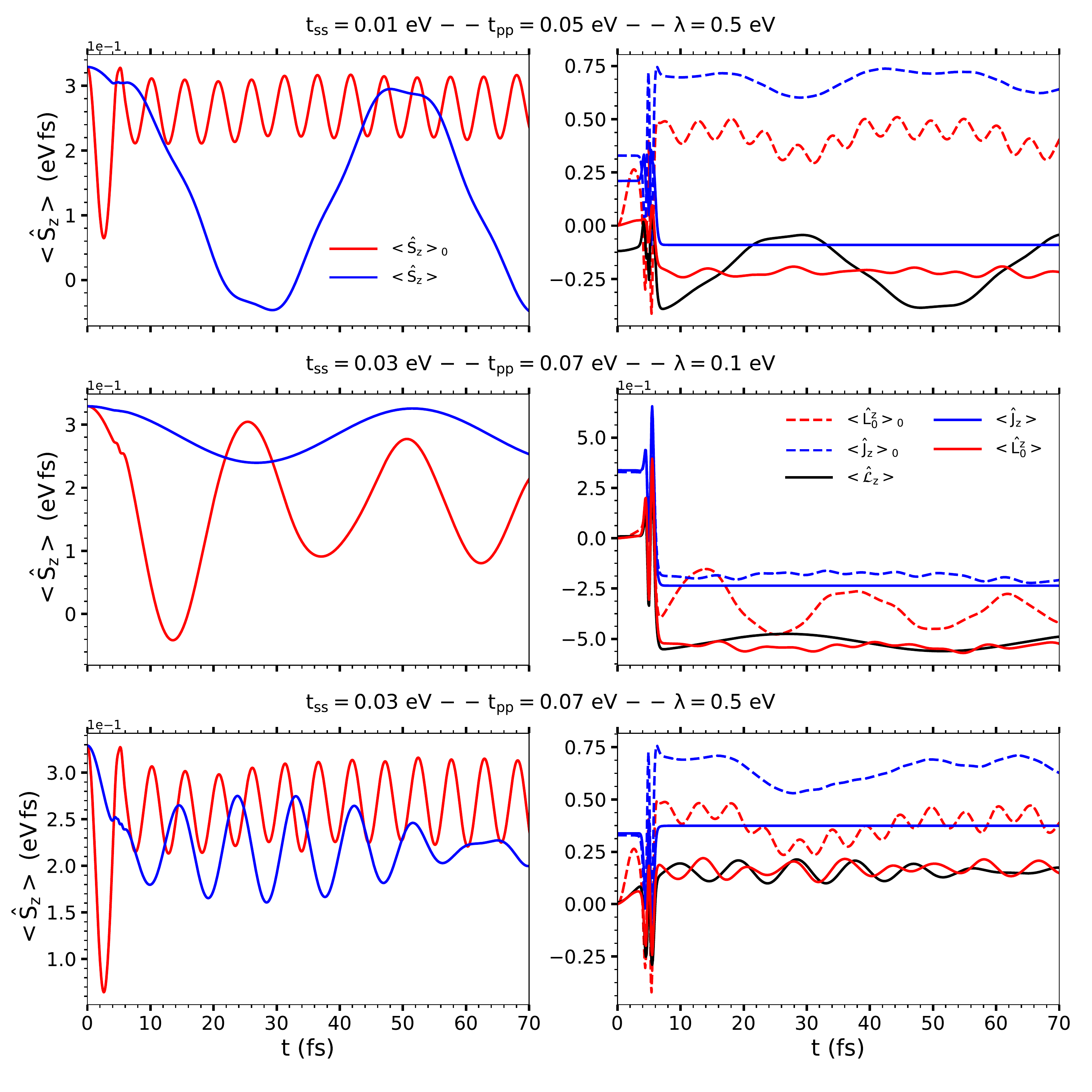}{(Color online) Time evolution of different observables for the atomic dimer excited by the application of an external laser field $E_x(t)$ slightly tilted with respect to the bond axis, with the effect of the atomic motion being neglected. In the upper plots we use a spin orbit strength $\lambda=0.5$~eV and hopping parameters $t_{ss}=0.01$~eV, $t_{pp}=0.05$~eV, we consider the expectation values $<\ldots>$ obtained by evolving the system's Hamiltonian with spin-orbit coupling from Eq.~(\ref{HSO_eq}) and $<\ldots>_0$ obtained by using the Hamiltonian with approximated spin-orbit from Eq.~(\ref{HSO0_eq}). In the plot on the left-hand side we compare the expectation value of the spin, $\hat{S}_z$, for the two calculations, while in the plots on the right-hand side we compare instead the expectation values of $\hat{L}_z^0$, $\hat{J}_z$ for the two calculations with $<\hat{\mathbfcal{L}}_z>$. For the middle plots we perform the same calculations with parameters $\lambda=0.1$~eV and $t_{ss}=0.03$~eV, $t_{pp}=0.07$~eV, while in the lower plots we use $\lambda=0.5$~eV and $t_{ss}=0.03$~eV, $t_{pp}=0.07$~eV.}{Fig:03}
The two spin-orbit operators considered here are
\begin{align}
&\hat{H}_{\rm SO}^0 = \sum_{a=1}^{N_i}\lambda_a \hat{\bf L}_a^0\cdot\hat{\bf S}\,,\label{HSO0_eq}\\
&\hat{H}_{\rm SO} = \sum_{a=1}^{N_i}\lambda_a \hat{\mathbfcal{L}}_a\cdot\hat{\bf S}\,.\label{HSO_eq}
\end{align}
Note that the first is of the type commonly used in tight-binding calculations and similar to that employed in rtTDDFT, while the second represents its generalization, as discussed in the previous section.
In Fig.~(\ref{Fig:03}) we evolve the electronic system under the action of the Hamiltonian (\ref{Eq:hamilt}), by approximating the spin-orbit coupling operator with Eq.~(\ref{HSO0_eq}) or with the complete spin-orbit coupling of Eq.~(\ref{HSO_eq}). The different sets of parameters used in the calculations are provided in the figure's caption, and we assume the atoms to be fixed in their initial positions. We distinguish the evolution of the observables obtained with spin-orbit coupling (\ref{HSO_eq}) from the ones obtained with spin-orbit (\ref{HSO0_eq}) by using respectively $<\ldots>$ and $<\ldots>_0$ for the expectation values.

As expected $<\hat{J}_z>=<\hat{\mathbfcal{L}}_z>+<\hat{S}_z>$ is a constant of motion after that the external electric-field pulse has vanished (see the solid blue line in the three plots on the right hand side of Fig.~(\ref{Fig:03}) after $t=10\,fs$), while $\expval*{\hat{J}_z}_0=\expval*{\hat{L}_z^0}_0+\expval*{\hat{S}_z}_0$ is not (see dotted blue lines). been set to zero (see solid blue line in the three plots on the right after $t=10\,fs$), while $<\hat{J}_z>_0=<\hat{L}_0^z>_0+<\hat{S}_z>_0$ is not (see dotted blue lines). Thus, the total orbital momentum is, in general, not conserved under the Hamiltonian including the approximated spin-orbit coupling interaction. The expectation value of the total orbital momentum $\expval*{\hat{L}_z^0}$ is also not conserved for both types of evolution (see red dotted and solid lines). The dynamics of $\expval*{\hat{L}_z^0}_0$ is oscillatory with the frequency that depends on the strength of the spin-orbit interaction, while the dynamics of $<\hat{L}_0^z>$ appears to be not as strongly dependent on the spin-orbit interaction.

The plots on the left hand side of Fig.~(\ref{Fig:03}) show the temporal evolution of the two observables $\expval*{\hat{S}_z}$ and $\expval*{\hat{S}_z}_0$. Their evolution is, not surprisingly, quite different, due to the fact that the operator $\hat{\mathbfcal{L}}$ is dependent on the strength of the hopping terms, since $[\hat{H}_{\rm ie}^0,\hat{\mathbfcal{L}}]=0$ must always be satisfied. The spin decay appears more pronounced for stronger spin-orbit interaction, although, for the same value of $\lambda$ it seems that the effect of the hopping terms should not be neglected either.
\myFig{1}{1}{true}{0}{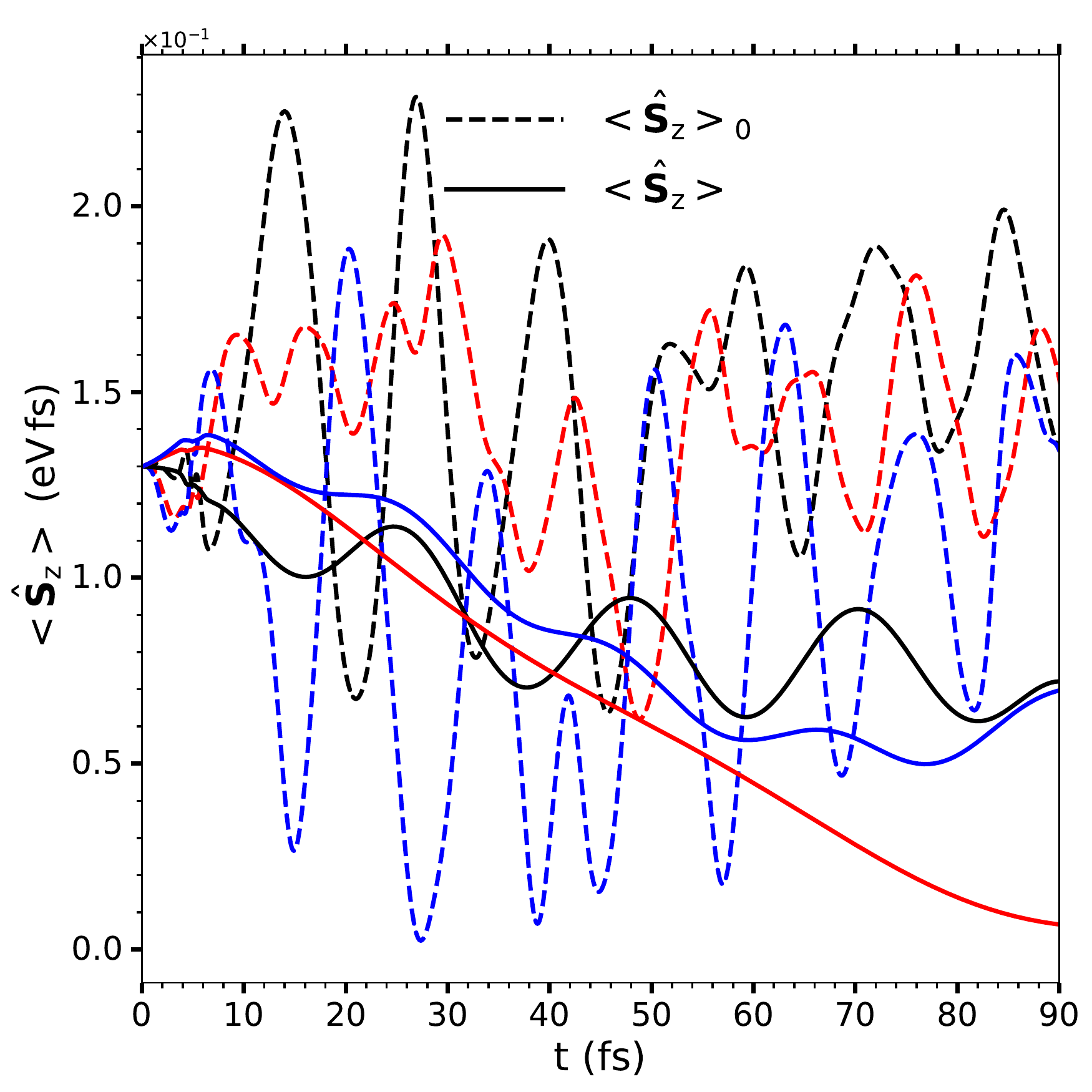}{(Color online) Time evolution of the observables $\expval*{\hat{S}_z}$ (solid lines) and $\expval*{\hat{S}_z}_0$ (dashed lines) for the different models of spin-orbit coupling in the case of a three-atoms system. The spin-orbit strength is fixed to a value $\lambda=0.2$~eV, $t_{ss}=0.03$~eV while $t_{sp}\simeq t_{pp}$ have values $\simeq\,0.12$~eV (black curves), $\simeq\,0.22$~eV (red curves) and $\simeq\,0.32$~eV (blue curves). The external electric field is applied along one of the bond axes.}{Fig:04}
In Fig.~(\ref{Fig:04}) we study the time evolution of a three-atom system initiated by an externally applied laser pulse and driven by two different models of spin-orbit coupling. The figure shows the evolution of the spin expectation values $\expval*{\hat{S}_z}$ and $\expval*{\hat{S}_z}_0$ when the spin-orbit coupling strength is fixed and the hopping terms are set to the values given in the caption. This choice of the parameters has no particular meaning from a physical point of view, and our purpose is to show how the choice may affect the orbital momentum operator $\hat{\mathbfcal{L}}$ and in turn the spin-orbit coupling, while its strength $\lambda$ is kept fixed. The evolution under the two spin-orbit operators is clearly different, in particular we observe that the non approximated one [Eq.~(\ref{HSO_eq})] leads to a greater loss of spin magnetization compared to the results obtained with the approximated evolution [Eq.~(\ref{HSO0_eq})]. The effect in the case of real magnetic system is still to be investigated, but already at this level it is clear that the two spin-orbit operators may lead to different rates of spin decay thanks to the different nature of the orbital operators $\hat{\mathbfcal{L}}$ and $\hat{\bf L}_0$. The fact that $\hat{\mathbfcal{L}}$ describes in addition to the localized orbital properties also the orbital properties of the delocalized electronic states leads to an additional contribution to the spin-orbit operator with a non trivial influence on the dynamical evolution of the magnetic system. 
\subsection{Coupling between electrons and molecular vibrations}
In this section we briefly consider the effect of molecular vibrations on the spin dynamics. In order to account for this effect our tight-binding Hamiltonian is modified to write
\begin{align}
&\hat{\tilde{H}}_{\rm ie}^{[{\bf R}]}(t) = \sum_{a=1}^{N_i}\sum_i\epsilon_{ai}\hat{c}_{ai}^\dagger\hat{c}_{ai} - \sum_{a<b}\sum_{i<j}\big[t_{ab}^{ij}({\bf R})\hat{c}_{bj}^\dagger\hat{c}_{ai}+{\rm h.c.}\big]\nonumber\\
&-\sum_{a<b}\delta{\bf R}(t)\cdot\sum_{i<j}\big[\boldsymbol{\nabla}t_{ab}^{ij}({\bf R})\hat{c}_{bj}^\dagger\hat{c}_{ai}+{\rm h.c.}\big] + \hat{H}_{\rm SO}^{[{\bf R}]} +\nonumber\\
&+\hat{H}_{\rm SO}^{[{\bf R}](1)} + \hat{\bf D}\cdot{\bf E}(t)\,,
\end{align}
where all the terms have the usual meaning and in addition $\delta{\bf R}_a(t)$ indicates the atomic vibration associated to atom $a$, $\boldsymbol{\nabla}t_{ab}^{ij}({\bf R})$ represents the transition matrix elements of the gradient of the electron-ion Hamiltonian $\hat{H}_{\rm ie}^0$ and $\hat{H}_{\rm SO}^{[{\bf R}](1)}$ is an additional contribution to the spin orbit coupling of Eq.~(\ref{HSO_eq}) arising when the atomic vibrations carry orbital momentum. From Eq.~(\ref{Eq:soc}) it is straightforward to show that
\begin{equation}
\hat{H}_{\rm SO}^a = -\frac{\partial_r\tilde{v}_{\rm ie}(r)}{2m_e^2 c^2|\hat{\bf r}-\hat{\bf R}_a|}\hat{\bf S}\cdot\big[\hat{\mathbfcal{L}}_a^0 + \delta\hat{\bf R}_a\times\hat{\bf P}_a\big]
\end{equation}
with the second term on the right hand side corresponding to the spin-vibration contribution $\hat{H}_{\rm SO}^{[{\bf R}](1)}$, while $\hat{\mathbfcal{L}}_a^0=({\bf R}_a^0-\hat{\bf r})\times\hat{\bf P}_a$ is the orbital momentum in the absence of atomic vibrations.
\myFig{1}{1}{true}{0}{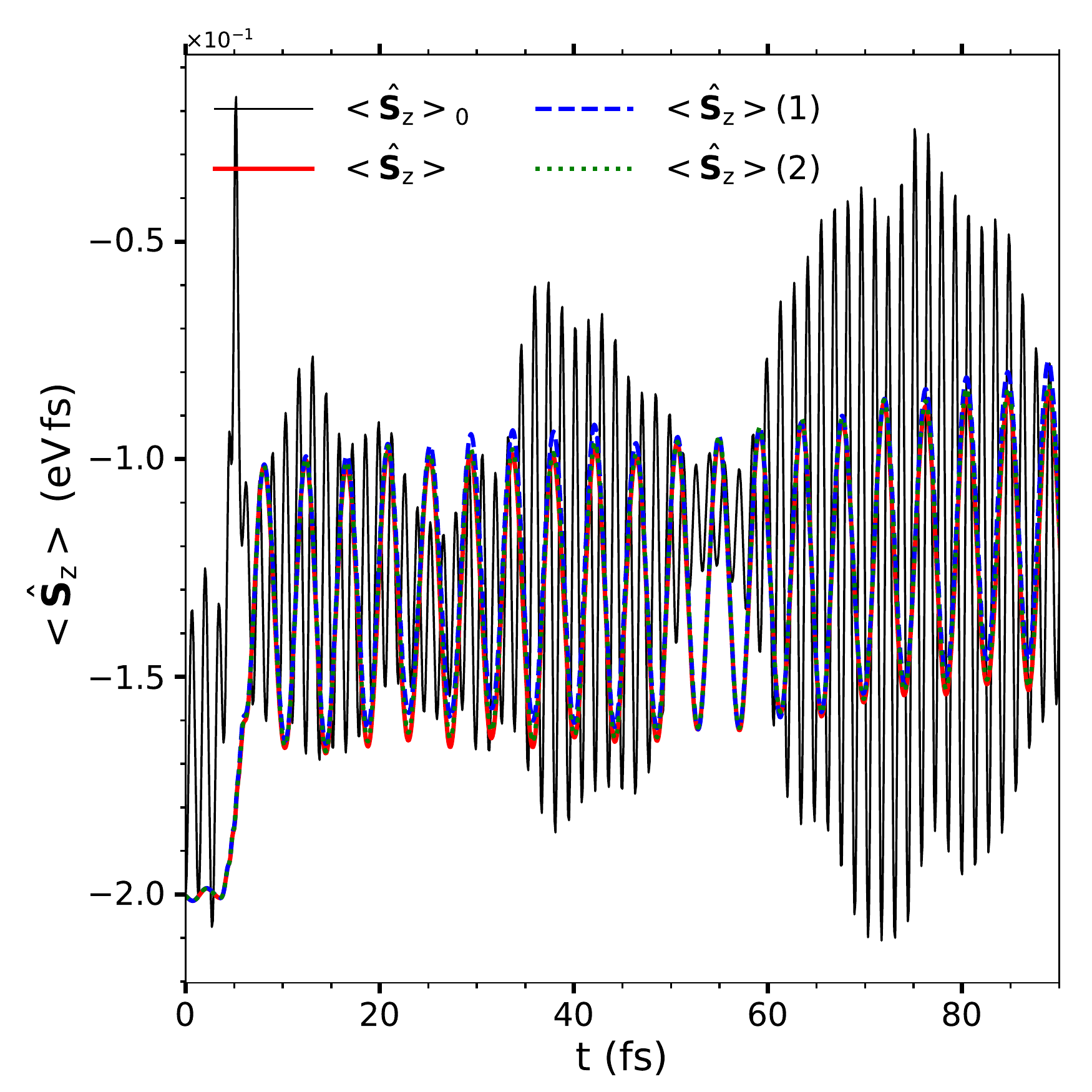}{(Color online) Time evolution of the observables $\expval*{\hat{S}_z}$ (solid red lines) and $\expval*{\hat{S}_z}_0$ (solid black lines) for the different models of spin-orbit coupling and without considering molecular vibrations in the case of a five-atoms system. The other two lines represent $\expval*{\hat{S}_z}$ obtained by using two different amplitudes of atomic vibrations, namely 0.03~\AA\ [green dotted line (2)] and 0.1~\AA\ [blue dashed line (1)].}{Fig:05}
In Fig.~(\ref{Fig:05}) we analyze the effect of these additional terms to the overall spin dynamics of a magnetic cluster of $5$ atoms excited by the application of an external laser pulse (the cluster has a pyramidal geometry with 4 atoms in plane and an additional atom out of plane). The vibrational orbital momentum acts also as an external field and its effect on the spin is shown in the figure for two different amplitudes of the vibrations $0.1\,\AA$ (blue dashed line) and $0.03\,\AA$ (green dotted line) and compared with the evolution in absence of vibrations for the exact (red line) and the approximated (black line) spin-orbit interaction. The effect does not seem particularly relevant at these time scales with only a noticeable change in the amplitude of the oscillations, while the evolution under the approximated spin-orbit interaction leads to huge modifications in the spin dynamics, as we have already observed.
\section{Conclusions and future works}\label{sec:5}
In this work we have discussed how to properly account for the conservation of the total orbital momentum in state-of-the-art \emph{ab initio} simulations of ultrafast demagnetization processes. We have identified the main problem behind the observed breaking of the full rotational invariance to be associated with the standard implementation of the spin-orbit coupling operator.
We have then explained why an approximated spin-orbit coupling operator, built as a function of the isolated atom orbital momentum, $\hat{\bf L}^0$, may lead to difficulties in the estimation of the spin and orbital momentum temporal evolution. A proper analysis of the ultrafast demagnetization process should not be limited to the correct description of the electron-atom (phonon) channel of energy and orbital momentum dissipation, but should be based also on a more accurate description of the spin-orbit coupling interaction.
In section \ref{sec:4} we have shown how, even for simple models, an oversimplified description of the spin-orbit coupling may affect the temporal evolution of the spin magnetization. These results are preliminary and based on simple model tight-binding Hamiltonians. At this point they only aim to show the importance of these new effects. In the future the model will be extended to describe more realistic systems (bulk materials and transition metal atomic clusters). This new description of the ultrafast spin dynamics has some interesting consequences and it may explain the apparent non-conservation of the orbital momentum observed in experiments\cite{Hennecke_2019}. The measure of the electronic orbital momentum is, in fact, performed by means of X-ray magnetic circular dichroism (XMCD), which provides separate information on the dynamics of the spin and the electronic orbital momentum. However, it is the expectation value of $\hat{\bf L}^0$, which is not a globally conserved quantity, to be directly accessible through these methods. In transition metals, $\hat{\mathbfcal{L}}_a$, is expected to significantly differ from $\hat{\bf L}_a^0$ around each atom, suggesting that the missing orbital momentum observed in experiments is not necessarily transfered to the nuclei but could remain hidden in the form of delocalized electronic orbital momentum.
\section{Acknowledgments}
JS was supported by the Molecular Foundry, a DOE Office of Science User Facility supported by the Office of Science of the U.S. Department of Energy under Contract No. DE-AC02-05CH11231. SS thanks the Irish Research Council (IRCLA/2019/127) for financial support.

\appendix

\section{Proof of Eq.~(\ref{Eq:Ltconserv})} \label{app:1}
We take Eq.~(\ref{Eq:Lconserv}) and we sum over all the ions $a$ on the left and the right hand side.
\begin{align}
\frac{d}{dt}\sum_{a=1}^{N_i}\hat{\mathbfcal{L}}_a &=-\frac{1}{m_e}\sum_{i=1}^{N_e}\hat{\bf p}_i\times\sum_{a=1}^{N_i}\hat{\bf P}_a+\sum_{i=1}^{N_e}\hat{\bf r}_i\times\sum_{a=1}^{N_i}\nabla_{{\bf R}_a}\hat{V}_{\rm ii}-\nonumber\\
&-\sum_{a=1}^{N_i}\hat{\bf R}_a\times\nabla_{{\bf R}_a}\hat{V}_{\rm ii}\,,
\end{align}
The first term on the right hand side is exactly zero due to the conservation of the linear momentum, $\sum_{i=1}^{N_e}\hat{\bf p}_i=-\sum_{a=1}^{N_i}\hat{\bf P}_a$. The second term is also zero. In Gaussian units we write
\begin{align}
&\sum_{i=1}^{N_e}\hat{\bf r}_i\times\sum_{a=1}^{N_i}\nabla_{{\bf R}_a}\hat{V}_{\rm ii}=\sum_{i=1}^{N_e}\hat{\bf r}_i\times\sum_{a=1}^{N_i}\sum_{a'\neq a}\nabla_{{\bf R}_a}\frac{(Ze)^2}{|{\bf R}_a - {\bf R}_{a'}|}\,\nonumber\\
&=\sum_{i=1}^{N_e}\hat{\bf r}_i\times\sum_{a=1}^{N_i}\sum_{a\neq a'}\bigg[-\frac{(Ze)^2\hat{{\bf R}}_a}{|{\bf R}_a-{\bf R}_{a'}|}+\frac{(Ze)^2\hat{\bf R}_{a'}}{|{\bf R}_a-{\bf R}_{a'}|}\bigg]\,\nonumber\\
&= 0\,.
\end{align}
Analogously for the third term we have
\begin{align}
&\sum_{a=1}^{N_i}\hat{\bf R}_a\times\nabla_{{\bf R}_a}\hat{V}_{\rm ii}=\sum_{a=1}^{N_i}\sum_{a'\neq a}(Ze)^2\frac{\hat{\bf R}_a\times\hat{\bf R}_{a'}}{|{\bf R}_a-{\bf R}_{a'}|^3}\nonumber\\
&=\frac{(Ze)^2}{2}\sum_{a=1}^{N_i}\sum_{a'\neq a}\bigg[\frac{\hat{\bf R}_a\times\hat{\bf R}_{a'}}{|{\bf R}_a-{\bf R}_{a'}|^3}-\frac{\hat{\bf R}_a\times\hat{\bf R}_{a'}}{|{\bf R}_a-{\bf R}_{a'}|^3}\bigg]\nonumber\\
&= 0\,,
\end{align}
that finally proves Eq.~(\ref{Eq:Ltconserv}), namely, that the operator $\sum_{a=1}^{N_i}\hat{\mathbfcal{L}}_a$ is a constant of motion for the system's Hamiltonian $\hat{H}$.
\begin{figure}[htbp]
  \begin{center}
  \begin{tikzpicture}
    \draw (0,0) -- (8,0) -- (8,4) -- (0,4) -- (0,0);
    \draw [-latex,dashed] (0,2) -- (8,2) node [above left]  {$x$};
    \node[circle,inner sep=1pt,fill=black,label=below:{$(0,0)$}] at (4,2) {};
    \draw [-latex,dashed] (4,0) -- (4,4) node [below right] {$y$};
    \draw[very thick] (2,2) -- (6,2);
    \shade [inner color=blue, outer color=white, even odd rule] (2,2) circle (0.45) circle (0.16);
    \draw[blue,thick,fill=blue] (2,2) circle (0.16cm);
    \node[circle,inner sep=1pt,fill=black,label={[label distance=0.1cm]30:$(-d/2,0)$}] at (2,2) {};
    \shade [inner color=blue, outer color=white, even odd rule] (6,2) circle (0.45) circle (0.16);
    \draw[blue,thick,fill=blue] (6,2) circle (0.16cm);
    \node[circle,inner sep=1pt,fill=black,label={[label distance=0.1cm]30:$(d/2,0)$}] at (6,2) {};
    \shade [inner color=blue, outer color=white, even odd rule] (5.732050807568878,3) circle (0.45) circle (0.16);
    \draw[blue,fill=blue] (5.732050807568878,3) circle (0.16cm);
    \node[circle,inner sep=1pt,fill=black,label={[label distance=0.1cm]30:$R_{1}$}] at (5.732050807568878,3) {};
    \draw[very thick,dashed] (4,2) -- (5.732050807568878,3);
    \shade [inner color=blue, outer color=white, even odd rule] (2.2679491924311224,1) circle (0.45) circle (0.16);
    \draw[blue,fill=blue] (2.2679491924311224,1) circle (0.16cm);
    \node[circle,inner sep=1pt,fill=black,label={[label distance=0.1cm]120:$R_{2}$}] at (2.2679491924311224,1) {};
    \draw[very thick,dashed] (4,2) -- (2.2679491924311224,1);
    \draw (5,2) arc (1.9106836:30:1) node [above] {$\delta\theta$};
  \end{tikzpicture}
  \end{center}
  \caption{(Color online) picture of the homonuclear dimer that we are using as a model system for the analysis of the orbital momentum conservation. In its initial configuration the bonding axis is along directed along $x$, in its successive configuration the entire dimer is rotated by an infinitesimal angle $\delta\theta$ around the out-of-plane $z$ axis. } \label{pic:dimer}
\end{figure}
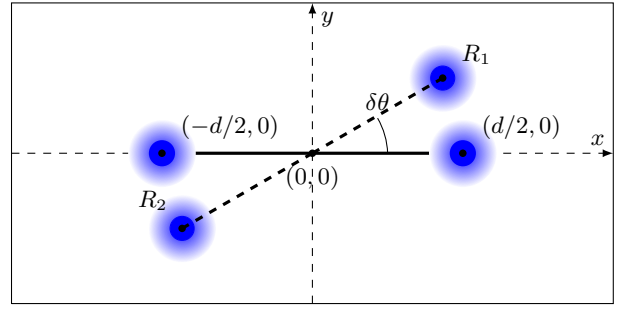
\section{Proof of inequality (\ref{TDTB})} \label{app:B}
Here we explicitly verify the inequality (\ref{TDTB}) for the electronic orbital momentum. The simplest possible case we may consider here is that of a homonuclear diatomic molecule (see Fig.~(\ref{pic:dimer}), the nature of the atoms is not relevant), with wave functions expanded over a minimal basis of $s$, $p_x$, $p_y$ and $p_z$ orbitals with no spin polarization. Here we do not want to describe a realistic situation, just prove conceptually that the orbital momentum of the atomic plus electronic subsystem cannot be conserved in general if we employ the operators $\hat{\bf L}_a^0$ to describe the electronic orbital momentum. According to Fig.~(\ref{pic:dimer}) we assume that the dimer is uniformly rotating around the $z$ axis (out-of plane) by an infinitesimal angle $\delta\theta$, the length of the bond is $d$, the Hamiltonian and orbital momentum operators are expanded over the atomic basis set $\{\ket{1;s},\ket{1;p_{-1}},\ket{1;p_0},\ket{1;p_1},\ket{2;s},\ket{2;p_{-1}},$\\
$\ket{2;p_0},\ket{2;p_1}\}$, in the new rotated configuration we can write  
\begin{equation}
\hat{L}_z^0 =
\begin{pmatrix}
  \Huge\mathbb{L}_{1,z}^0
  & \rvline & \bigzero \\
\hline
  \bigzero & \rvline &
  \Huge\mathbb{L}_{2,z}^0
\end{pmatrix}
=
\begin{bmatrix}
0 & 0 & 0 & 0 & 0 & 0 & 0 & 0\\
0 &-1 & 0 & 0 & 0 & 0 & 0 & 0\\
0 & 0 & 0 & 0 & 0 & 0 & 0 & 0\\
0 & 0 & 0 & 1 & 0 & 0 & 0 & 0\\
0 & 0 & 0 & 0 & 0 & 0 & 0 & 0\\
0 & 0 & 0 & 0 & 0 &-1 & 0 & 0\\
0 & 0 & 0 & 0 & 0 & 0 & 0 & 0\\
0 & 0 & 0 & 0 & 0 & 0 & 0 & 1\\
\end{bmatrix}
\end{equation}
%
that corresponds to the total electronic orbital momentum operator, obtained as a direct sum of the orbital momentum operators corresponding to the two isolated atomic sites, $\hat{\bf L}^0 = \hat{\bf L}_1^0\oplus\hat{\bf L}_2^0$. The hamiltonian $\hat{H}_{\rm ie}$ of the system, in turn, depends on $4$ free parameters, the on-site orbital energies $\epsilon_s$, $\epsilon_p$ (we assume degeneracy of the $p$ orbitals for simplicity) and the hopping coefficients depend on a set of primitive integrals $t_{ss}$, $t_{sp}$, $t_{pp}^\sigma$ and $t_{pp}^\pi$ that here are left arbitrary but in general are a function of the distance between the two atoms of the dimer molecule. These integrals $t_{ij}^{nlm_a;n'l'm'_a}$ are diagonal in the angular momentum projection $m_a$ along the bond axis and do not depend on its sign\cite{Koskinen_2009}, $t_{ij}^{nlm_a;n'l'm'_a}=t_{ij}^{nl|m_a|;n'l'|m_a|}\delta_{m_a,m_a'}$. These can be written through the following general expression
\begin{equation}
t_{ij}^{nlm_a,n'l'm_a}=\int_Vd^3{\bf r}\psi_{nlm_a}^*({\bf r}-{\bf R}_i)h_{ij}({\bf r})\psi_{n'l'm_a}({\bf r}-{\bf R}_j),
\end{equation}
where $\psi_{nlm_a}$ is an atomic wave function and $h_{ij}({\bf r})=-\hbar^2\boldsymbol{\nabla}^2/2m_e+v_{\rm ie}({\bf r}-{\bf R}_i)+v_{\rm ie}({\bf r}-{\bf R}_j)$, the free hopping parameters are given by the following two-centers integrals
\begin{eqnarray}
t_{ss} & = & \mel{n=2,l=0,m_a=0}{\hat{h}_{12}}{n=2,l'=0,m_a=0}\,,\nonumber\\
t_{sp} & = & \mel{n=2,l=0,m_a=0}{\hat{h}_{12}}{n=2,l'=1,m_a=0}\,,\nonumber\\
t_{pp}^\sigma & = & \mel{n=2,l=1,m_a=0}{\hat{h}_{12}}{n=2,l'=1,m_a=0}\,,\nonumber\\
t_{pp}^\pi & = & \mel{n=2,l=1,m_a=1}{\hat{h}_{12}}{n=2,l'=1,m_a=1}\,.\nonumber
\end{eqnarray}
In addition we also have the exact relation $t_{ij}^{nlm_a,n'l'm_a}=(-1)^{l+l'}t_{ji}^{nlm_a,n'l'm_a}$ that leads to the final expression for the dimer's Hamiltonian
\begin{widetext}
$$
\small{\hat{H}_{\rm ie}=}
\begin{bsmallmatrix}
\epsilon_s & 0 & 0 & 0 & t_{ss} & \frac{\lambda_-t_{sp}}{\sqrt{2}} & t_{sp}\lambda_z & -\frac{\lambda_+t_{sp}}{\sqrt{2}}\\
0 & \epsilon_p & 0 & 0 & -\frac{\lambda_-t_{sp}}{\sqrt{2}} & \frac{t_{pp}^\sigma+t_{pp}^\pi-\lambda_z^2\Delta t_{pp}}{2} & \frac{\lambda_z\lambda_-\Delta t_{pp}}{\sqrt{2}} & -\frac{\lambda_-^2\Delta t_{pp}}{2} \\
0 & 0 & \epsilon_p & 0 & -t_{sp}\lambda_z& \frac{\lambda_z\lambda_-\Delta t_{pp}}{\sqrt{2}} & \lambda_z^2 \Delta t_{pp}+t_{pp}^\pi & -\frac{\lambda_z\lambda_+\Delta t_{pp}}{\sqrt{2}} \\
0 & 0 & 0 & \epsilon_p & \frac{\lambda_+t_{sp}}{\sqrt{2}} & -\frac{\lambda_-^2\Delta t_{pp}}{2} & -\frac{\lambda_z\lambda_+\Delta t_{pp}}{\sqrt{2}} & \frac{t_{pp}^\sigma+t_{pp}^\pi-\lambda_z^2\Delta t_{pp}}{2}\\
t_{ss} & -\frac{\lambda_+}{\sqrt{2}}t_{sp} & -t_{sp}\lambda_z & \frac{\lambda_-}{\sqrt{2}}t_{sp} & \epsilon_s & 0 & 0 & 0\\
\frac{\lambda_+}{\sqrt{2}}t_{sp} & \frac{t_{pp}^\sigma+t_{pp}^\pi-\lambda_z^2\Delta t_{pp}}{2} & \frac{\lambda_z\lambda_+}{\sqrt{2}}\Delta t_{pp} & -\frac{\lambda_-^2}{2}\Delta t_{pp} & 0 & \epsilon_p & 0 & 0\\
 t_{sp}\lambda_z & \frac{\lambda_z\lambda_+}{\sqrt{2}}\Delta t_{pp} & \lambda_z^2\Delta t_{pp}+t_{pp}^\pi & -\frac{\lambda_z\lambda_-}{\sqrt{2}}\Delta t_{pp} & 0 & 0 & \epsilon_p & 0\\
 -\frac{\lambda_-}{\sqrt{2}}t_{sp}& -\frac{\lambda_+^2}{2}\Delta t_{pp} & -\frac{\lambda_z\lambda_-}{\sqrt{2}}\Delta t_{pp} & \frac{t_{pp}^\sigma+t_{pp}^\pi-\lambda_z^2\Delta t_{pp}}{2} & 0 & 0 & 0 & \epsilon_p\\
\end{bsmallmatrix}
$$
\end{widetext}
where we have used $\Delta t_{pp}=t_{pp}^\sigma-t_{pp}^\pi$, the dimer bond length is d=$|{\bf R}_1-{\bf R}_2|$ with $\lambda_{x,y,z}=({\bf R}_1-{\bf R}_2)_{x,y,z}/|{\bf R}_1-{\bf R}_2|$ and $\lambda_\pm=\lambda_x\pm i\lambda_y$. 

In the rotated configuration we have $({\bf R}_1-{\bf R}_2)_x=d\cos\delta\theta$, $({\bf R}_1-{\bf R}_2)_y=d\sin\delta\theta$ and we can substitute in the Hamiltonian
\begin{eqnarray}
\lambda_\pm & = & 1 -\frac{\delta\theta^2}{2} \pm i\delta\theta\,,\\
\lambda_z & = & 0\,,
\end{eqnarray}
the new atomic coordinatets according to Fig.~(\ref{pic:dimer}) is given by ${\bf R}_1=\frac{d}{2}(\cos\delta\theta,\sin\delta\theta,0)$, ${\bf R}_2=-\frac{d}{2}(\cos\delta\theta,\sin\delta\theta,0)$. The time derivative of the orbital momentum of the two atoms is given by
\begin{eqnarray}
\dot{\bf L}_1 & = & -{\bf R}_1\times\boldsymbol{\nabla}_{{\bf R}_1}\expval*{\delta\hat{H}_{\rm ie}} = -{\bf R}_1\times\boldsymbol{\nabla}_{{\bf R}_1}\delta\theta\cdot\partial_{\delta\theta}\expval*{\delta\hat{H}_{\rm ie}}\nonumber\,\\
\dot{\bf L}_2 & = & -{\bf R}_2\times\boldsymbol{\nabla}_{{\bf R}_2}\expval*{\delta\hat{H}_{\rm ie}} = -{\bf R}_2\times\boldsymbol{\nabla}_{{\bf R}_2}\delta\theta\cdot\partial_{\delta\theta}\expval*{\delta\hat{H}_{\rm ie}}\nonumber\,
\end{eqnarray}
and the total orbital momentum time derivative becomes
\begin{align} \label{Eq:Ldot}
\dot{\bf L}^{\rm atom} &= \dot{\bf L}_1 + \dot{\bf L}_2 = -\partial_{\delta\theta}\expval*{\hat{H}_{\rm ie}}\cdot({\bf R}_1-{\bf R}_2)\times\boldsymbol{\nabla}_{{\bf R}_1}\delta\theta\,,\nonumber\\
&=-4\hat{\bf e}_z\cdot\partial_{\delta\theta}\expval*{\hat{H}_{\rm ie}}\,.
\end{align}
In order to simplify the calculation we proceed by assuming that we have a single electron in a state given by the superposition of $\ket{1;s}$ around atom $1$ and $\ket{2;p_{-1}}$ around atom $2$; $\ket{\Psi}=[1,0,0,0,0,1,0,0]/\sqrt{2}$ in the same basis of the Hamiltonian. By explicitly computing Eq.~(\ref{Eq:Ldot}) we obtain at first order in $\delta\theta$
\begin{align}
&\partial_{\delta\theta}\expval*{\hat{H}_{\rm ie}} = -\frac{t_{sp}}{\sqrt{2}}\delta\theta\,,\nonumber\\
&\dot{\bf L}^{\rm atom} = 2\sqrt{2}\cdot\hat{\bf e}_z t_{sp}\delta\theta\,.
\end{align}
The time derivative of the expectation value of the electronic orbital momentum is given by ($\hbar$ is set to $1$)
\begin{equation}
\frac{d}{dt}\expval*{\hat{L}_z^0}=i\expval*{[\hat{H}_{\rm ie},\hat{L}_z^0]}\,,
\end{equation}
that explicitly computed by using the wave function $\ket{\Psi}$ previously defined gives
\begin{equation}
\frac{d}{dt}\expval*{\hat{L}_z^0} = - \frac{t_{sp}}{\sqrt{2}}\delta\theta\,.
\end{equation}
The time derivative of the total orbital momentum $\dot{\bf L}^{\rm atom}+\frac{d}{dt}\expval*{\hat{L}_z^0}$ along the $z$ axis is
\begin{equation}
\dot{L}_z^{\rm atom} + \frac{d}{dt}\expval*{\hat{L}_z^0} = \frac{3}{\sqrt{2}}t_{sp}\delta\theta\,,
\end{equation}
that proves Eq.~(\ref{Eq:Ltconserv}) in the main text.

\end{document}